\documentclass[11pt]{article}

\def\showauthornotes{0}

\usepackage{amsfonts}
\usepackage{amssymb}
\usepackage{amstext}
\usepackage{amsthm}
\usepackage{bm}
\usepackage{color}
\usepackage{comment} 
\usepackage{enumitem}
\usepackage{fancybox}
\usepackage{hyperref}
\usepackage{ifthen}
\usepackage{subfigure}
\usepackage{tabularx}
\usepackage{textcomp}
\usepackage{thm-restate}
\usepackage{thmtools}
\usepackage{url}
\usepackage{verbatim}
\usepackage{xcolor}
\usepackage{mathtools}

\newtheorem{theorem}{Theorem}[section]

\newtheorem{lemma}[theorem]{Lemma}

\newtheorem{conjecture}[theorem]{Conjecture}

\newtheorem{fact}[theorem]{Fact}

\theoremstyle{definition}
\newtheorem{definition}[theorem]{Definition}

\DeclareMathOperator*{\pr}{\mathbb{P}}
\DeclareMathOperator*{\av}{\mathbb{E}}
\newcommand{\tr}[1]{\mathop{\mbox{Tr}}\left({#1}\right)}

\newcommand\rea{\mathbb R}

\newcommand{\ol}[1]{\ensuremath{\overline{#1}}}
 
\newcommand\calA{\mathcal{A}}

\newcommand\calV{\mathcal{V}}
\newcommand\calW{\mathcal{W}}
\definecolor{Mygray}{gray}{0.8}

 \ifcsname ifcommentflag\endcsname\else
  \expandafter\let\csname ifcommentflag\expandafter\endcsname
                  \csname iffalse\endcsname
\fi

\ifnum\showauthornotes=1
\newcommand{\todo}[1]{\colorbox{Mygray}{\color{red}#1}}
\else
\newcommand{\todo}[1]{}
\fi

\ifnum\showauthornotes=1
\newcommand{\Authornote}[2]{{\sf\small\color{red}{[#1: #2]}}}
\newcommand{\Authoredit}[2]{{\sf\small\color{red}{[#1]}\color{blue}{#2}}}
\newcommand{\Authorcomment}[2]{{\sf \small\color{gray}{[#1: #2]}}}
\newcommand{\Authorfnote}[2]{\footnote{\color{red}{#1: #2}}}
\newcommand{\Authorfixme}[1]{\Authornote{#1}{\textbf{??}}}
\newcommand{\Authormarginmark}[1]{\marginpar{\textcolor{red}{\fbox{%
#1:!}}}}
\else
\newcommand{\Authornote}[2]{}
\newcommand{\Authoredit}[2]{}
\newcommand{\Authorcomment}[2]{}
\newcommand{\Authorfnote}[2]{}
\newcommand{\Authorfixme}[1]{}
\newcommand{\Authormarginmark}[1]{}
\fi

\newcommand{\paren}[1]{\left({#1}\right)}
\newcommand{\sqparen}[1]{\left[{#1}\right]}
\newcommand{\curlyparen}[1]{\left\{{#1}\right\}}
\def\pleq{\preccurlyeq}
\def\pgeq{\succcurlyeq}

\def\abs#1{\left|#1  \right|}

\def\norm#1{\left\| #1 \right\|}

\newcommand\PPi{\boldsymbol{\Pi}}

\newcommand\bb{\boldsymbol{\mathit{b}}}

\newcommand\ee{\boldsymbol{\mathit{e}}}

\newcommand\rr{\boldsymbol{\mathit{r}}}

\newcommand\uu{\boldsymbol{\mathit{u}}}
\newcommand\vv{\boldsymbol{\mathit{v}}}
\newcommand\ww{\boldsymbol{\mathit{w}}}
\newcommand\yy{\boldsymbol{\mathit{y}}}
\newcommand\zz{\boldsymbol{\mathit{z}}}
\newcommand\xx{\boldsymbol{\mathit{x}}}

\renewcommand\AA{\boldsymbol{\mathit{A}}}
\newcommand\BB{\boldsymbol{\mathit{B}}}
\newcommand\CC{\boldsymbol{\mathit{C}}}
\newcommand\DD{\boldsymbol{\mathit{D}}}

\newcommand\II{\boldsymbol{\mathit{I}}}

\newcommand\LL{\boldsymbol{\mathit{L}}}
\newcommand\PP{\boldsymbol{\mathit{P}}}

\newcommand\XX{\boldsymbol{\mathit{X}}}

\newcommand\Otil{\widetilde{O}}
\newenvironment{tight_enumerate}{
\begin{enumerate}
 \setlength{\itemsep}{2pt}
 \setlength{\parskip}{1pt}
}{\end{enumerate}}

\newcommand{\vone}{\boldsymbol{\mathbf{1}}}
\newcommand{\vzero}{\boldsymbol{\mathbf{0}}}

\usepackage{xcolor}
\usepackage{cleveref}
\usepackage{pdfsync}
\usepackage{authblk}

\usepackage[backend=biber, isbn=false, style=alphabetic, backref=true, doi=false, url=false, maxcitenames=10, mincitenames=5, maxalphanames=10, maxbibnames=10, minbibnames=5, minalphanames=5, defernumbers=true, sortlocale=en_US]{biblatex}

\usepackage[margin=1in]{geometry}
\usepackage[linesnumbered,noend,boxed,noline]{algorithm2e}
\DontPrintSemicolon
\IncMargin{1pt}
\SetFuncSty{textsc}

\SetKwProg{Pro}{Procedure}{}{}
    \SetKwFunction{CS}{CycleSparsify}
    \SetKwFunction{CSO}{CycleSparsifyOnce}
    \SetKwFunction{CD}{CycleDecomposition}
    \SetKwFunction{PCS}{ColourSparsify}
    \SetKwFunction{PCSO}{ColourSparsifyOnce}
    \SetKwFunction{CW}{ColourWeights}
    \SetKwFunction{ED}{ExpanderDecomp}
    \SetKwFunction{SS}{SampleSparsify}
    \SetKwFunction{SSO}{SampleSparsifyOnce}
    \SetKwFunction{PD}{PatchDegrees}
    \SetKwFunction{CO}{CorrectOrientation}
    \SetKwFunction{CT}{ColourTarget}
    \SetKwFunction{ACT}{AlternativeColourTarget}
    \SetKwFunction{PC}{PartialColour}
    \SetKwFunction{PCG}{ColourSparsifyGraph}
    \SetKwFunction{PCC}{ColourSparsifyCycle}

\newcommand{\Znote}{\Authornote{Z}}

\ifnum\showauthornotes=1
\newcommand{\anvith}[1]{\textcolor{brown}{anvith: #1}}
\else
\newcommand{\anvith}[1]{}
\fi

\crefname{fact}{fact}{facts}
\Crefname{fact}{Fact}{Facts}
\Crefname{algocf}{Algorithm}{Algorithms}
\Crefname{conjecture}{Conjecture}{Conjectures}

\DeclareMathOperator{\lev}{lev}
\DeclareMathOperator{\im}{im}
\DeclareMathOperator{\reff}{Reff}

\DeclareMathOperator{\rev}{rev}

\newcommand\Ltil{\widetilde{\boldsymbol{\mathit{L}}}}

\newcommand\tcd{\textnormal{T}_{\textnormal{CD}}}

\DeclareMathOperator{\hlift}{\mathsf{hlift}}

\begin{document}
\title{Better Sparsifiers for Directed Eulerian Graphs}
\author{Sushant Sachdeva
}
\affil{
University of Toronto\\
\texttt{sachdeva@cs.toronto.edu}
}
\author{Anvith Thudi}
\affil{
University of Toronto\\
\texttt{anvith.thudi@mail.utoronto.ca}
}
\author{Yibin Zhao}
\affil{
University of Toronto\\
\texttt{yibin.zhao@cs.toronto.edu}
}

\maketitle
\thispagestyle{empty}

\begin{abstract}
  Spectral sparsification for directed Eulerian graphs is a key
  component in the design of fast algorithms for solving directed
  Laplacian linear systems.
  Directed Laplacian linear system solvers are crucial algorithmic primitives to
  fast computation of fundamental problems on random walks, such as
  computing stationary distribution, hitting and commute time, and personalized
  PageRank vectors.
  While spectral sparsification is well understood for undirected
  graphs and it is known that for every graph $G,$
  $(1+\varepsilon)$-sparsifiers with $O(n\varepsilon^{-2})$ edges
  exist [Batson-Spielman-Srivastava, STOC '09] (which is optimal), the
  best known constructions of Eulerian sparsifiers require
  $\Omega(n\varepsilon^{-2}\log^4 n)$ edges and are based on
  short-cycle decompositions [Chu et al., FOCS '18].

  In this paper, we give improved constructions of Eulerian
  sparsifiers, specifically:
  \begin{tight_enumerate}
      \sloppy
    \item We show that for every directed Eulerian graph $\vec{G},$ there exist
      an Eulerian sparsifier with
      $O(n\varepsilon^{-2} \log^2 n \log^2\log n +
      n\varepsilon^{-4/3}\log^{8/3} n)$ edges.  This result is based on
      combining short-cycle decompositions
      [Chu-Gao-Peng-Sachdeva-Sawlani-Wang, FOCS '18, SICOMP] and
      [Parter-Yogev, ICALP '19], with recent progress on the matrix
      Spencer conjecture [Bansal-Meka-Jiang, STOC '23].
    \item We give an improved analysis of the constructions based on
          short-cycle decompositions,
    giving an $m^{1+\delta}$-time
    algorithm for any constant $\delta > 0$ for constructing Eulerian sparsifiers with
    $O(n\varepsilon^{-2}\log^3 n)$ edges.
  \end{tight_enumerate}
\end{abstract}

\newpage
\setcounter{page}{1}
\section{Introduction} 
\label{sec:intro}
Given a graph $G(V, E),$ a sparsifier of $G$ is a graph $H$ on the
same set of vertices $V,$ but hopefully supported on a subset of the edges
$E' \subset E$ such that $H$ approximately preserves certain
properties of $G.$
Several notions of graph sparsification have been well studied for
undirected graphs, e.g. spanners (approximately preserving distances),
cut sparsifiers, spectral sparsifiers, etc.

Spectral sparsification is a particularly influential notion of undirected graph
sparsification~\cite{SpielmanT04}. 
Spectral sparsifiers generalize cut-sparsifiers introduced by
Benczur-Karger~\cite{BenczurK96}, which guaranteed that the total
weight of every vertex cut is preserved up to a multiplicative factor
of $(1+\varepsilon)$ in the sparsifier.
Efficient spectral sparsification was one of the core developments
that led to the development of nearly-linear time solvers for
Laplacian linear systems~\cite{SpielmanT04}. It further inspired the
Laplacian paradigm, resulting in faster algorithms for many graph
problems including
sampling/counting random spanning
trees~\cite{DurfeeKPRS17, DurfeePPR17},
approximating edge centrality measures~\cite{LiZ18} etc.

The first construction of spectral sparsifiers for undirected graphs
by Spielman and Teng required
$\Omega( n \varepsilon^{-2} \text{poly}(\log n) )$ number of edges
with a large, unspecified power of $\log n.$ Subsequently, Spielman
and Srivastava~\cite{SpielmanS08:journal} gave a very simple and elegant
construction, whereby sampling each edge independently with a
probability proportional to its leverage score results in a spectral
sparsifier with $O(n\varepsilon^{-2} \log n)$ edges with high
probability. In a complete graph, sampling edges independently with
probability $p$ requires $p = \Omega(\varepsilon^{-2}\log n)$ to
achieve $(1+\varepsilon)$-spectral sparsification; thus demonstrating that
such a construction requires $\Omega(n\varepsilon^{-2} \log n)$ edges.

Batson-Spielman-Srivastava~\cite{BatsonSS12} further improved this to
show that there exist spectral sparsifiers for undirected graphs with
$O(n\varepsilon^{-2})$ edges, and that we cannot do better even for
the complete graph. Thus, they essentially settled the question of
optimal size of undirected spectal sparsifiers.

For directed graphs, sparsification has been trickier to define. It is
immediate to see that in a complete bi-partite graph with all edges
directed from the left vertices to the right vertices, if one wishes
to approximately preserve all directed cuts, one must preserve all the
edges. This means that there is no non-trivial cut-sparsification (or
its generalization) for arbitrary directed graphs.

Such pathological cases can be avoided if one restricts to Eulerian
directed graphs, i.e. a graph where each vertex has its total weighted
in-degree equal to its total weighted out-degree, in which case cut sparsification
becomes equivalent to cut sparsification of undirected graphs.
Indeed, Cohen-Kelner-Peebles-Peng-Rao-Sidford-Vladu~\cite{CKPPRSV17}
show that it is possible to define a meaningful generalization of
spectral sparsification (and hence cut sparsification) to Eulerian
directed graphs. We will call these sparsifiers Eulerian sparsifiers
for brevity. In a manner similar to the original Spielman-Teng
construction, they give a nearly-linear time $\Otil(m)$-time algorithm
to build an Eulerian sparsifier with
$O(n\varepsilon^{-2}\text{poly}(\log n))$ edges, with a large
unspecified power of $\log n.$

Since Eulerian sparsification generalizes undirected spectral
sparsification, $\Omega(n\varepsilon^{-2})$ edges are necessary for
constructing Eulerian sparsifiers.  There has been some progress on
proving the existence of Eulerian sparsifiers with fewer edges:
Chu-Gao-Peng-Sachdeva-Sawlani-Wang~\cite{CGPSSW18} introduced the
notion of short-cycle decomposition, a decomposition of an unweighted
graph as a union of short edge-disjoint cycles, and a few extra
edges. As a simple lemma, they show that every undirected graph can be
represented as a union of edge-disjoint cycles of length $2 \log n,$
with at most $2n$ extra edges.
Using this short-cycle decomposition,~\cite{CGPSSW18} were able to
prove that Eulerian sparsifiers with $O(n\varepsilon^{-2}\log^4 n)$
edges exist. However, the following natural question remains
unanswered:
\begin{quote}
  What is the best possible sparsity guarantee for constructing
  Eulerian sparsifiers?
\end{quote}

In this paper, we make progress on this question.  First, we present
an improved analysis of the short-cycle based Eulerian-sparsification
from~\cite{CGPSSW18}.
\begin{restatable}{theorem}{informalcycle} \label{thm:informalcycle}
  For every constant $\delta > 0,$ there is an algorithm that taken as
  input a directed Eulerian graph $\vec{G}$ and returns an
  $\varepsilon$-Eulerian sparsifier of $\vec{G}$ with
  $O(n\varepsilon^{-2}\log^3 n)$ edges in $m^{1+\delta}$ time.
  \end{restatable}
The above algorithm is based on independently toggling short cycles,
i.e., with probability 1/2, all the clockwise edges are deleted and
counter-clockwise edges are doubled, and with probability 1/2 the
the counter-clockwise edges are deleted and the clockwise edges
doubled.
Given that the edges in each $O(\log n)$ length short-cycle are
toggled in a completely correlated manner, and the cycles are toggled
independently, this approach naturally cannot lead to a sparsity
better than $O(n\varepsilon^{-2} \log^2 n).$ 

To go beyond the above result, we leverage discrepancy theory, and
specifically the recent progress on the matrix Spencer conjecture by
Bansal, Jiang, and Meka~\cite{BJM23}.
(Please see the related works section, \Cref{subsec:related}, for a description
of the matrix Spencer conjecture.)
While the matrix Spencer conjecture itself is not directly useful for our
application, we are able to utilize the underlying machinery from~\cite{BJM23}
towards the conjecture together with the short-cycle decomposition to prove the
following:
\begin{theorem}[Informal] \label{thm:informalexist} There is an
  algorithm that given a Eulerian graph $\vec{G},$ can compute in
  poly-time a $\varepsilon$-Eulerian sparsifier of $\vec{G}$ with
  $n \varepsilon^{-2} \log^2 n + n \varepsilon^{-3/4} \log^{8/3} n$
  edges (up to $\log \log n$ factors).
\end{theorem}
For small $\varepsilon$, e.g. $\varepsilon^{-1} = \Omega(\log n),$ the
above theorem gives an $n \varepsilon^{-2} \log^2 n$ bound, only a
$\log^2 n$ factor away from the lower bound.

\subsection{Related works} \label{subsec:related}
\paragraph{Sparsification.}
In the realm of undirected spectral sparsification, there are four
major approaches: expander decomposition \cite{SS11,ACKQWZ16,JS18},
spanners \cite{KP12,KX16, KPPS17}, importance sampling
\cite{SS08,KLP12}, and potential function based sparsification
\cite{BSS14,ALO15,LS15,LS17}. 

More closely related to Eulerian
spectral sparsification is undirected degree preserving
sparsification, introduced by \cite{CGPSSW18}.  Degree preserving
sparsification is useful for constructing spectral sketches.  More
importantly for us, 
techniques for degree preserving sparsification can generally be extended to
work for directed Eulerian sparsification.
The standard notion of Eulerian approximation (and sparsification), 
first introduced by \cite{CKPPRSV17},
requires exact preservation of the differences between in and out degrees while
ensuring the difference in directed Eulerian Laplacians is small with
respect to the Laplacian of the undirectification of the graph.
That is, for $\epsilon \in (0,1)$,
\[
    \norm{\LL_G^{+/2} (\LL_{\vec{H}} - \LL_{\vec{G}}) \LL_G^{+/2}} \leq
    \epsilon.
\]
This definition subsumes degree preserving sparsification for undirected graphs.
However, the difficulty in Eulerian sparsification compared to degree preserving is that
the directed Eulerian Laplacians are neither symmetric nor positive
semidefinite while undirected Laplacians satisfy both.

\cite{CKPPRSV17} showed the first degree preserving (implicitly) and
Eulerian sparsifier using expander decomposition.
The algorithm performs random sampling of the directed edges with probability related to the degrees within each expander.
A small patching is then added at the end to fix up the degrees for each
expander.
Recent work on different notions of Eulerian sparsification \cite{APPSV23}
establishes an ``equivalence'', 
albeit with significantly stronger requirements
than spectral approximation,
between degree preserving and Eulerian sparsification under the notion
of singular value approximation.  They established the first Eulerian
sparsifier with both nearly-linear sparsity and nearly-linear runtime
using this connection. However, the best known efficient constructions of
sparsifiers based on expander decompositions lose a large
$\text{poly}(\log n)$ factor~\cite{APPSV23}.  Even ignoring efficient
construction, the expander approach has a natural lower bound of at
least a $\log^2 n$ factor in the worst-case sparsity even for the
undirected spectral sparsification due to a lowerbound on the optimal
tradeoff between expansion factor and number of expanders (see
\cite{SW19}).

As an alternative to using expander decompositions, the technique of using short
cycles for sparsification introduced by \cite{CGPSSW18} also applies to both
degree preserving and Eulerian sparsifications with sparsity $O(n\epsilon^{-2}
\log^2 n)$ and $O(n\epsilon^{-2} \log^4 n)$ respectively.
Improved short cycle decompositions were subsequently designed in~\cite{LSY19,
PY19} to facilitate faster construction of sparsifiers.
Our first result \Cref{thm:informalcycle} follows closely
to \cite{CGPSSW18} and reduces the gap between degree-preserving and
Eulerian sparsification under this technique. 

Recently \cite{JRT23} demonstrated a new approach in degree preserving
sparsification using discrepancy theory.
They showed that the operator norm discrepancy bodies are 
well conditioned 
(i.e., satisfy certain Gaussian measure lowerbound) 
for symmetric and positive semidefinite matrices that arise from undirected
sparsification and used an approximate version of the framework from \cite{RR23}
to give a colouring of the edges (corresponding to adding and deleting edges)
under the linear constraint needed for degree preservation.
However, %
the underlying discrepancy bodies studied by \cite{JRT23}
do not align well with Eulerian sparsifications
where matrices are no longer positive semidefinite 
and matrix variance statistic is the primary statistic one has control over (see
\Cref{sec:existential}).

\paragraph{Directed Laplacian solvers.}
\cite{CKPPSV16} initiated the line of work that studies the problem of solving
directed Laplacian linear systems.
They established a reduction from solving general directed Laplacian systems to
Eulerian Laplacian systems and motivated subsequent studies in Eulerian
Laplacian solvers.
\cite{CKPPRSV17} gave an almost linear time algorithm for solving Eulerian
Laplacians using the squaring identities from \cite{PengS14}.
\cite{CKKPPRS18} gave the first nearly linear time solver using the standard
approximate LU factorization techniques that enjoyed great success in undirected
Laplacian solvers \cite{KLPSS16,SZ23}.
\Znote{Perron-Frob}
\cite{AhmadinejadJSS19} further established a reduction from solving linear
systems of (asymmetric) M-matrices to Eulerian Laplacian systems, giving fast
comuptation of several problems closely associated with the Perron-Frobenius
theorem.
\cite{PS22} extended the approach from \cite{CKKPPRS18} and gave an
approach for extending an algorithm for building Eulerian sparsifiers
to a fast solver for Eulerian Laplacian linear systems.
Combined with \Cref{thm:informalcycle}, they give a $O(n\log^4 n
\log(n/\epsilon))$ time solver with $m^{1+\delta}$ preprocessing time for any
constant $\delta > 0$.
\cite{RasmusMG22} established the first derandomized directed Laplacian solver
in almost linear time.

\paragraph{Discrepancy theory.}
The Matrix Spencer Conjecture \cite{Zou12,Mek14} is a major open problem in
discrepancy theory:
\begin{conjecture}[Matrix Spencer Conjecture] \label{conj:spencer}
    \sloppy
    Given $n \times n$ symmetric matrices $\AA_1, \ldots, \AA_m \in
    \mathbb{R}^{n \times n}$ with $\| \AA_i \| \leq 1$, there exist signs
    $x \in \{ \pm 1 \}^m$ such that $\| \sum_{i=1}^m x_i \AA_i \| \leq
    O(\sqrt{m} \cdot \max\{1,\sqrt{\min\{1,\log(n/m)\}}\})$.
\end{conjecture}
As a natural comparison, for an uniform random colouring $x \in \{ \pm 1\}^m$, 
the matrix Chernoff bound \cite{Tropp12} gives that
\[
    \av \sqparen{\norm{ \sum_i x_i \AA_i }} = 
    O \left(\sqrt{\log n} \right) \cdot \norm{ \sum_i \AA_i^2 }^{1/2}
    \leq O(\sqrt{m \log n})
\]
which has a gap of $\sqrt{\log n}$ to \Cref{conj:spencer} when $m \geq n$.
We refer readers to \cite{LRR17,HRS22,DJR22} for recent progress toward solving
this conjecture.

Many natural problems in studying the spectra of matrices can be viewed as
matrix discrepancy theory problems, e.g., graph sparsification
\cite{BSS14,RR20} and the
Kadison-Singer problem \cite{MSS15}.
\cite{RR20} studies the geometry of operator norm balls for a
collection of matrices where, $\| \sum_i | \AA_i | \|$ is small.
This result was subsequently used in \cite{JRT23} to show optimal degree preserving
sparsification. 
As previously mentioned, this line of work is not applicable for the purpose of
Eulerian sparsification since matrices that emerge from our setting do not
satisfy that $\| \sum_i | \AA_i | \|$ is small.
\cite{BJM23} resolved the Matrix Spencer Conjecture for matrices of rank
$n /(\log^{O(1)} n)$ using a recent advancement in matrix concentration
bounds due to \cite{BBvH23}.
The partial colouring result for controlling operator norm used in \cite{BJM23}
serves as the main machinery in our existential results (see
\Cref{lemma:colouring}).
Specifically, the matrices we study naturally satisfy $\| \sum_i \AA_i^2 \|$ is
small.

\subsection{Technical overview}

Our approach to constructing Eulerian sparsifiers builds on the framework
introduced in \cite{CGPSSW18}.
For completeness, the entire algorithm and our improved analysis are
presented in \Cref{sec:cycle}. 

The sparsification algorithm in \cite{CGPSSW18} combines importance
sampling of edges with a short cycle decomposition.
At each iteration, the algorithm restricts its attention to edges with
small ``importance'' in the undirected graph (edges with leverage
score $w_e \bb_e^\top \LL_G^+ \bb_e$ at most constant times the
average leverage score, $O(n/m)$).
Note that there are $\Omega(m)$ such edges. The algorithm then
performs a short cycle decomposition on these edges -- expressing the
graph as a union of uniformly weighted edge-disjoint short cycles and a few
extra edges.
For each short cycle, the algorithm independently keeps either the clockwise edges
or the counter-clockwise edges with probability ${1}/{2}$ each.  The
number of edges reduces by a constant fraction overall in expectation
at each iteration.  After doubling the weights of the cycle edges
retained, the algorithm guarantees that the Eulerianess of each short
cycle is preserved and, hence, the entire graph.  Moreover, when
combined with the undirected leverage score condition above, such
changes in directed short cycles also have a small variance overall.
The matrix Bernstein inequality for asymmetric matrices guarantees a
small approximation error for this randomized step.  We repeat this
process until the desired approximation error is met.

To obtain our improved analysis of this algorithm, we first present
improved variance bounds for random matrices corresponding to short
cycles (see \Cref{lemma:cyclefpart}).  Rather than bounding the
variance terms through complete graphs as in \cite{CGPSSW18}, we
instead bound it directly with respect to the undirected cycle.  This
improved analysis also serves a critical role in our partial colouring
approach presented in \Cref{sec:existential}.

In the rest of our paper, we present our existential result which uses
the partial colouring lemma from \cite{BJM23} (stated in
\Cref{lemma:colouring}) to choose how to sparsify the short cycles.
\begin{restatable}{lemma}{colouring}[\cite{BJM23} Lemma 3.1] 
    \label{lemma:colouring}
    There exists constants $c,c' > 0$ such that the following holds.
    Given symmetric matrices $\AA_1, \ldots, \AA_m \in \rea^{n \times n}$ that
    satisfies $\| \sum_{i=1}^m \AA_i^2 \| \leq \sigma^2$ and 
    $\sum_{i=1}^m \| \AA_i \|_F^2 \leq mf^2$ and a point $\yy \in (-1,1)^m$, 
    there is an algorithm \PC that returns
    a point $\xx \in [-1,1]^m$ such that
    \begin{equation} \label{eq:colour}
        \norm{\sum_{i=1}^m (x_i - y_i) \AA_i} \leq 
        c(\sigma + (\log^{3/4} n) \sqrt{\sigma f}),
    \end{equation}
    and $|\{i : x_i \in \{\pm 1\}\}| > c' m$ in polynomial time.
\end{restatable}

Our improved algorithm will follow the same high-level approach as the random
sampling construction discussed above.  For each directed short cycle,
instead of independently randomly choosing to preserve either the
clockwise or the counter-clockwise edges, we will use the partial
coloring $\xx$ given by the \Cref{lemma:colouring}. In each iteration, the
\Cref{lemma:colouring} allows the algorithm to compute a partial colouring with
sufficiently many fully coloured entries (i.e., entries with value
$\pm 1$) on all the short cycles.  This colouring indicates which part
to remove.  The partial colouring lemma allows us to remove these
parts of the cycles with less error than random sampling.

However, there are two major challenges in applying \Cref{lemma:colouring}.
Firstly, within each iteration, we cannot afford to fully colour all the cycles
by recursively applying \Cref{lemma:colouring}, since
in the worst case, we need to perform the partial colouring $O(\log n)$
times, which will result in an additional $\log$ factor in the sparsity.
Hence, we are always left with non-fully coloured cycles (i.e., entries with
magnitude $<1$).
The cycles that are non-fully coloured must still be incorporated into the
sparsified graph after each iteration to guarantee the error given by the
partial colouring lemma.
However, we cannot explicitly modify the graph to include edges corresponding to
these cycles.
If we were to do so, we would lose the integral and polynomially bounded weight
conditions and the short cycle decomposition could no longer be applied to this
new graph for further sparsification.

On the other hand, naively incorporating partially coloured cycles into the next
iteration is also problematic.
Unlike the undirected case, the two parts of a directed cycle do not necessarily
have the same number of edges.
For example, if a directed cycle has all edges in the same direction, then one
part has all the edges and the other part none.
If we start our colouring process from a non-zero initial partial
colouring (i.e., a non-zero $\yy$ to \Cref{lemma:colouring}), we could end up at
a colouring where almost no edges are removed.

To deal with these problems, our algorithm handles the integral weighted portion
$\vec{G}$ of the graph $\vec{G}'$ and the non-fully coloured cycles $\ol{S}$
separately (see \Cref{alg:coloursparse}).
For the integral weighted portion, we perform the partial colouring to guarantee
at least a constant fraction of edges are removed. 
We then add the non-fully coloured cycles into the set $\ol{S}$.
For the set of non-fully coloured cycles $\ol{S}$, we carefully adjust their
colouring by considering the difference between the partial colours and $\pm
1$.
We ensure that a good portion of cycles in $\ol{S}$ are fully coloured after the
procedure to guarantee the size of $\ol{S}$ does not blow up.
In both cases, the approximation error incurred by the partial colouring
operation is precisely controlled to guarantee our desired final error and hence
\Cref{thm:informalexist}.

\section{Preliminaries} 
\label{sec:background}

\paragraph{General notation.}
We use $\Otil(\cdot)$ to suppress polylog factors in $n,m$.
We say ``with high probability in $n$'' to mean an event suceeeds with
probability $1 - n^{-\Omega(1)}$ for an arbitrary constant. 
In the context of graphs, $n$ is assumed to be the number of vertices and is often
omitted.
All logarithms throughout the paper are with base 2.

\paragraph{Linear Algebra.}
We use boldface to denote vectors.
The all-zeros and all-ones vector are $\vzero$ and $\vone$.
For any $u$, we let $\ee_u$ to denote the vector such that the $u$th coordinate
is $1$ and all other coordinates are 0.
We denote $\bb_{uv} = \ee_u - \ee_v$ for any $u \neq v$.
For vectors $\uu,\vv$ of equal dimension, $\uu \circ \vv$ is the entrywise
product.
For a linear subspace $\calW$ of a vector space $\calV$, we denote $\calW^\perp$
as the orthogonal complement of $\calW$ in $\calV$.

Matrices are denoted in boldface capticals. 
We use $\ker(\AA),\im(\AA)$ to denote the kernel and image of $\AA$.
For any $u$, we let $(\AA)_u$ denote the $u$th column of $\AA$.
A symmetric matrix $\AA$ is positive semidefinite (PSD)
(resp. positive definite (PD)) if, for any vector
$\xx$ of compatible dimension, $\xx^\top \AA \xx \geq 0$ (resp. $\xx^\top
\AA \xx > 0$ ).
Let $\AA$ and $\BB$ be two symmetric matrices of the same dimension,
then we write $\BB \pleq \AA$ or $\AA
\pgeq \BB$ if $\AA-\BB$ is PSD.
The ordering given by $\pleq$ is called Loewner partial order. 
\begin{fact} \label{fact:twoside}
    If $\AA \pgeq \BB$ and $\CC$ is any matrix of compatible dimension, then
   $ \CC \AA \CC^\top \pgeq \CC \BB \CC^\top.$
\end{fact}
Let $\norm{\AA}$ and $\norm{\AA}_F = \sqrt{\tr{\AA^* \AA}}$ denote the operator
norm and Frobenius norm of a matrix $\AA$.
The operator norm is equal to the largest singular value of $\AA.$
For a matrix $\AA \in \rea^{n \times m}$, we define the Hermitian (symmetric)
lift of $\AA$ by
\[
    \hlift(\AA) = 
    \begin{bmatrix}
        & \AA \\
        \AA^\top & 
    \end{bmatrix}
    \in \rea^{(n+m) \times (n+m)}
\]
The norms of Hermitian lifts satisfy $\norm{\hlift(\AA)} = \norm{\AA}$ and
$\norm{\hlift(\AA)}_F = 2\norm{\AA}_F$.
Given a symmetric matrix with eigenvalue decomposition $\AA = \sum_{i}
\lambda_i \vv_i \vv_i^{\top},$ where $\{\vv_i\}_i$ form an orthonormal
basis, the pseudoinverse is defined as 
$\AA^+ = \sum_{i: \lambda_i \neq 0} \frac{1}{\lambda_i}\vv_i \vv_i^{\top}.$
The absolute value of $\AA$ on eigenvalues is defined as
$|\AA| = \sum_{i: \lambda_i \neq 0} |\lambda_i| \vv_i \vv_i^{\top}$.
Note that $|\AA|$ is PSD.
Similarly for symmetric PSD matrix $\AA$ we have 
$\AA^{1/2} = \sum_{i: \lambda_i \neq 0} \sqrt{\lambda_i}\vv_i \vv_i^{\top}$ and
$\AA^{+/2} = \sum_{i: \lambda_i \neq 0} 
\frac{1}{\sqrt{\lambda_i}} \vv_i \vv_i^{\top}$.

\paragraph{Graphs and Laplacians.} 
Let $\vec{G} = (V,E,\ww)$ be a weighted directed graph (possibly with
multi-edges) with edge weights $\ww: E \rightarrow \rea_{\ge 0}$.
We write $G$ as the corresponding undirected graph of $\vec{G}$ where each
directed edge $e \in E(\vec{G})$ correspond to an undirected edge on the same
incident vertices with half its weight.
A weighted directed graph $\vec{G}$ is Eulerian if for each vertex $v \in V$,
its weighted in degree equals its weighted out degree.

We associate to a weighted directed graph $\vec{G}$ a matrix $\LL_{\vec{G}}
\in \rea^{V \times V}$ called the directed Laplacian of $\vec{G}$.
We define a non-negative matrix $\AA_{\vec{G}}$ as the adjacency matrix of
$\vec{G}$ with $\AA_{uv} = w_{uv}$ if $(u,v) \in E$ and $\AA_{uv} = 0$
otherwise.
The weighted degree matrix of $\vec{G}$ is a non-negative diagonal matrix
$\DD_{\vec{G}}$ corresponding to the weighted out-degrees of $\vec{G}$.
Then, $\LL_{\vec{G}} = \DD_{\vec{G}} - \AA_{\vec{G}}^\top$ and satisfies
$\vone^\top \LL_{\vec{G}} = \vzero^\top$, i.e. 
$(\LL_{\vec{G}})_{uu} = -\sum_{v \neq u} \LL_{vu}$ for all $u \in V$.
For a weighted Eulerian directed graph $\vec{G}$, its graph Laplacian
additionally satisfies $\LL_{\vec{G}} \vone = \vzero$.
Assuming Eulerian graph $\vec{G}$, the associated undirected graph Laplacian
matrix of $G$ is $\LL_G = \frac{1}{2}(\LL_{\vec{G}} + \LL_{\vec{G}}^\top)$.
$\LL_G$ is symmetric and PSD.
For an undirected Laplacian $\LL_G$, the effective resistance and leverage score
of an edge $e \in E(G)$ is defined by $\reff_G(e) = \bb_e^\top \LL_G^+ \bb_e$
and $\lev_G(e) = w_e \reff_G(e)$ where we fixed an arbitrary orientation for the
undirected edge $e$.

We assume $n$ and $m$ to be the number of vertices and edges in a graph.
As is standard, we consider strongly connected Eulerian graphs with positive
integral and polynomially bounded edge weights.
Whenever we say the weights are polynomially bounded, we assume they are bounded
by $n^{O(1)}$.

\section{Eulerian sparsification via short cycle decomposition} 
\label{sec:cycle}

We first present an improved analysis of constructing Eulerian sparsifiers using short cycle decompositions analogous to \cite{CGPSSW18}. In particular, we provide a better variance analysis of the error terms in sparsification than what was used by \cite{CGPSSW18}; by Matrix Bernstein~(Theorem~\ref{thm:bernstein}) this will allow us to use fewer edges to retain a desired error bound.

We first recall the definition of a short cycle decomposition of a graph $G$.

\begin{definition} \label{def:cycledecomp}
    An $(\hat{m},L)$-short cycle decomposition of an unweighted undirected graph
    $G$, decomposes $G$ into several edge-disjoint cycles, each of length at
    most $L$, and at most $\hat{m}$ edges are not in the union of the cycles.
\end{definition}

We assume $\CD$ is an algorithm that takes as input an unweighted graph with $n$
vertices and $m$ edges and returns a $(\hat{m},L)$-short cycle decomposition in
time $\tcd(m,n)$.
As in \cite{CGPSSW18}, we also assume the \emph{super-additivity} of $\tcd$:
\[
    \sum_i \tcd(m_i, n) \leq \tcd\left( \sum_i m_i, n \right)
\]
for all $m_i \geq n$.

Relevant to us is the following construction of a short cycle decomposition.

\begin{lemma}[\cite{PY19} Theorem 2]\label{lemma:cdfast}
    \sloppy
    For any $\delta > 0,$ there is an algorithm that computes an
    $(O(n\log n), O(2^{1/\delta} \log n))$-short cycle decomposition of an
    undirected unweighted graph in $2^{O(1/\delta)}m n^{\delta}$ time.
\end{lemma}
Thus, for every constant $\delta > 0,$ we have an $m^{1+\delta}$-time
algorithm that computes an $(O(n\log n),\\ O(\log n))$-short cycle
decomposition.

\begin{algorithm}[t]
    \caption{Make a cycle have a consistent orientation/direction}
    \label{alg:correct_orientation}
    \Pro{\CO{$\vec{C}$}}{
        Pick an arbitrary edge $e_1$ in $\vec{C}$ and let $v_1$ be its tail
        vertex. Define $V_{\vec{C}}$ as the vertex set of $\vec{C}$.\;
        Initialize $E_{\vec{S}} \leftarrow \emptyset$, $E_{\vec{F}} \rightarrow \{e_1\}$, $V_{\vec{F}} = \{ v_1\}$, $i = 1$ \;
        \While{$|V_{\vec{C}} \setminus V_{\vec{F}}| > 0$}{$i \leftarrow i+1$ \;
        Take $e_{i+1}$ be the other edge incident on $v_i$. \;
        If $e_{i+1}$ is outgoing from $v_i$, take $v_{i+1}$ the head of $e_{i+1}$ and update $E_{\vec{F}} \leftarrow E_{\vec{F}} \cup \{e_{i+1}\}$, $V_{\vec{F}} \leftarrow V_{\vec{F}} \cup \{v_{i+1}\}$. \;
        Else let $v_{i+1}$ be the tail of $e_{i+1}$ and update $E_{\vec{S}}
        \leftarrow E_{\vec{S}} \cup \{e_{i+1}\}$, $E_{\vec{F}} \leftarrow
        E_{\vec{F}} \cup \{\rev(e_{i+1})\}$, $V_{\vec{F}} \leftarrow V_{\vec{F}}
        \cup \{v_{i+1}\}$.\;
        }
        \Return{$\vec{F}$ defined by $E_{\vec{F}}$ and $V_{\vec{F}}$, and $S$
        the undirected graph defined by $E_{\vec{S}}$ and the incident vertices
        of $E_{\vec{S}}$.}\;
    }
\end{algorithm}

\begin{algorithm}[t]
    \caption{Sparsification via short cycle decomposition}
    \label{alg:cyclesparse}
    \Pro{\CS{$\vec{G}, \epsilon$, \CD}}{
        Decompose each edge by its binary representation.\;
        Compute $\rr$ a 1.5-approximate effective resistances in $G$.\;
        \While{$|E(\vec{G})| \geq O(\hat{m}\log n + \epsilon^{-2} n L^2 \log n)$}{
            $\vec{G} \gets$ \CSO{$\vec{G},\rr$,\CD}.\;
        }
        \Return{$\vec{G}$.}\;
    }
\end{algorithm}

\begin{algorithm}[t]
    \caption{Sparsify once using short cycles}
    \label{alg:cycleonce}
    \Pro{\CSO{$\vec{G},\rr,\CD$}}{
        \KwIn{A directed Eulerian graph $\vec{G}$ where edge weights are
            integral powers of 2,%
            a 2-approximate effective resistances $\rr$ in $G$, 
            a short cycle decomposition algorithm \CD.
        }
        \KwOut{A directed Eulerian graph $\vec{H}$ where edge weights are
            integral powers of 2.
        }
        \BlankLine
        $\vec{H} \gets \vec{G}$ with only the edges which satisfies 
        $w_e r_e > \frac{4n}{m}$ and remove these edges from $\vec{G}$.\;
        Partition $\vec{G}$ into uniformly weighted graph 
        $\vec{G}_1, \ldots, \vec{G}_s$ where $\vec{G}_i$ has all edge weights
        $2^i$ and $s = O(\log n)$.\;
        \For{each $\vec{G}_i$}{
            $\{ C_{i,1}, \ldots, C_{i,t} \} \gets$ \CD{$G_i$} and let
            $\vec{C}_{i,j}$ be the corresponding directed graph of $C_{i,j}$ in
            $\vec{G}_i$.\;
            $\vec{H} \gets \vec{H} + \vec{G}_i \setminus 
            \left( \bigcup_{j=1}^t \vec{C}_{i,j} \right)$.\;
            \For{each cycle $\vec{C}_{i,j}$}{
                With probability $1/2$, add all its clockwise edges with twice
                their weight to $\vec{H}$.
                Otherwise, add the counter-clockwise edges instead.\; 
                \label{alg:cycleonce_add}
            } 
        }
        \Return{$\vec{H}$.}\;
    }
\end{algorithm}

Our random sampling based sparsification algorithm is the same as
\cite{CGPSSW18}. 
We repeatedly sparsify an Eulerian graph by keeping only the ``clockwise" or
``counter-clockwise" edges of each cycle in a short cycle decomposition of the
graph, see \CS in \Cref{alg:cyclesparse} and \CSO in \Cref{alg:cycleonce}.

Stated in other words, we will sparsify a cycle by
partitioning it into two sets and removing one of those sets randomly.
For a directed cycle $\vec{C}$, we take $\vec{F},S$ to be the outputs of
\CO{$\vec{C}$}.
In particular, $\vec{F}$ is the cycle $\vec{C}$ corrected so that every vertex
has an incoming edge and an outgoing edge, and $S$ is the undirected graph
coming from the set of edges in $\vec{C}$ whose direction we reversed 
(where the edge weight in $S$ are the same as the original edge weights).
We consider the direction of edges defined by $\vec{F}$ as clockwise.
Then, the edges in $S$ are all the counter-clockwise edges in $\vec{C}$.

For a cycle $C$ and its corresponding directed cycle $\vec{C}$, the directed
graph Laplacian added at \autoref{alg:cycleonce_add} in \CSO is the following:
\[
    \begin{cases}
        \LL_{\vec{C}} + \LL_{\vec{F}} - \LL_{S} & \mbox{w.p. } 1/2 \\
        \LL_{\vec{C}} - \LL_{\vec{F}} + \LL_{S} & \mbox{w.p. } 1/2
    \end{cases}
\]
which means the changes incurred on the directed graph Laplacian is
\begin{equation} \label{eq:diff}
    \begin{cases}
        \Ltil  & \mbox{w.p. } 1/2 \\
        -\Ltil & \mbox{w.p. } 1/2
    \end{cases},
    \mbox{where } \Ltil = \LL_{\vec{F}} - \LL_{S}.
\end{equation}
Note that this change preserves the difference between the in and out degrees of
$\vec{C}$.
Either a vertex had an incoming and outgoing edge (and so difference $0$),
in which case both edges are either in $\vec{F} \setminus S$ or in $S$ and
hence always added together with the same weights (so still difference
$0$).
Alternatively a vertex had two incoming or outgoing edges, in which case
only one is ever added with twice the weight, which then still preserves the
difference between in and out degree.

To analyze the approximation error incurred by each round of sparsification, we
first recall a bound on the Laplacian of $S$.
\begin{lemma}[\cite{CGPSSW18} Lemma 5.5] \label{lemma:cyclespart}
    If $\vec{C}$ is an equal weighted directed cycle of length $L$ contained in a graph $\vec{G}$ where each edge
    $\vec{e} \in \vec{C}$ satisfies $\lev_G(e) \leq \rho$,
    then $\LL_S \preceq (L \cdot \rho) \cdot \LL_G$.
\end{lemma}

We will also need the following helper lemma which bounds the effect of
$L_{\vec{F}}$.
Compared to Lemma 5.6 in \cite{CGPSSW18}, our result improves the bound by a factor of
$L$.
\begin{lemma} \label{lemma:cyclefpart}
    If $\vec{C}$ is a equal weighted directed cycle of length $L$
    contained in a graph $\vec{G}$ where each edge
    $\vec{e} \in \vec{C}$ satisfies $\lev_G(e) \leq \rho$.
    Then, $\LL_{\vec{F}}^\top \LL_G^+ \LL_{\vec{F}} \preceq
    O(L^2\rho) \LL_C $.
\end{lemma}
\begin{proof}
    Let $\PPi_C = \II_C - \frac{1}{L}\vone_C \vone_C^\top$ be the projection
    matrix on the support of $C$ except the all one vector on $C$. 
    Notice that $\ker^\perp(\LL_C) = \im(\LL_C) = \im(\PPi_C)$.
    Furthermore, we have $\im(\LL_{\vec{F}}) \subset \im(\PPi_C)$
    (as $\vone_{C} \in \im^{\perp}(\LL_{\vec{F}})$) so $\PPi_C
    \LL_{\vec{F}} = \LL_{\vec{F}}$, and also
    $\im^{\perp}(\PPi_C) \subset \ker(\LL_{\vec{F}}^{\top})$ hence
    $\LL_{\vec{F}}^{\top} \PPi_C = \LL_{\vec{F}}^{\top}$.
    Thus, $\LL_{\vec{F}}^\top \LL_G^+ \LL_{\vec{F}}
    = \LL_{\vec{F}}^\top \PPi_C \LL_G^+ \PPi_C \LL_{\vec{F}}$.

    Let $w$ be the weight of each edge in $\vec{C}$. 
    Then, $\LL_{\vec{F}} = w(\II - \PP)$ where $\PP$ is a permutation matrix on
    the vertices of $C$ corresponding to the transition matrix $\vec{F}$
    and $\LL_C = \frac{w}{2}(2\II - \PP - \PP^\top)$.
    Now, $\LL_{\vec{F}}^\top \PPi_C
    \LL_{\vec{F}} = \LL_{\vec{F}}^\top \LL_{\vec{F}} =
    w^2(\II - \PP^\top)(\II-\PP) = w^2(2\II - \PP - \PP^\top) = 2w \LL_C$. 
    It then suffices to show $\PPi_C \LL_G^+ \PPi_C \preceq O(L^2 \rho/w) \PPi_C$. 
    As $\ker(\LL_G^+) \subseteq \ker(\PPi_C)$, it also suffices to show $\|
    \PPi_C \LL_G^+ \PPi_C \| = O(L^2 \rho/w)$.
    We can write out each entry of $\PPi_C$ by 
    $(\PPi_C)_u =
    \frac{1}{L-1}\sum_{v \in C, v \neq u} \bb_{uv}$ for $u \in C$ and 0 otherwise.
    As effective resistance is a metric, 
    $w \bb_{uv}^\top \LL_G^+ \bb_{uv} \leq (L-1) \rho$ for any distinct vertices
    $u,v \in C$.
    Note that this factor of $L$ is an upperbound on the combinatorial distance
    from $u$ to $v$ in $C$.
    Then,
    \begin{align*}
        &|(\PPi_C)_x^\top \LL_G^+ (\PPi_C)_u | 
        \\
        =&
        \abs{
            (\frac{1}{L-1}\sum_{y \in C, y \neq x} \LL_G^{+/2} \bb_{xy})^\top 
            (\frac{1}{L-1}\sum_{v \in C, v \neq u} \LL_G^{+/2} \bb_{uv})
        }
        \\
        \leq& 
        \sum_{y \neq x, y \in C} \sum_{v \neq u, v \in C} 
        \frac{1}{w^{1/2}(L-1)} \norm{ w^{1/2} \LL_G^{+/2} \bb_{xy} } \cdot 
        \frac{1}{w^{1/2}(L-1)} \norm{ w^{1/2} \LL_G^{+/2} \bb_{uv} }
        \\
        \leq& 
        (L-1)^2 \times \frac{(L-1)\rho}{wL^2} \leq L\rho/w.
    \end{align*}
    By Gershgorin circle theorem and the length of $C$, any eigenvalue of
    $\PPi_C \LL_G^{+} \PPi_C$ cannot exceed $L^2 \rho/w$ as required.
\end{proof}

With \Cref{lemma:cyclespart} and \Cref{lemma:cyclefpart} we can now bound a term
involving $\tilde{L}$ which will appear later in our variance analysis.

\begin{lemma} \label{lemma:cyclebounds}
    Let $\vec{C}$ is an equal weighted directed cycle of length $L$
    contained in a graph $\vec{G}$ where each edge
    $\vec{e} \in \vec{C}$ satisfies $\lev_G(e) \leq \rho$.
    Then $\LL_G^{+/2} (\Ltil^\top \LL_G^+ \Ltil) \LL_G^{+/2} \preceq O(L^2\rho)
    \cdot \LL_G^{+/2}\LL_C \LL_G^{+/2}$ and $\LL_G^{+/2} (\Ltil \LL_G^+
    \Ltil^\top) \LL_G^{+/2} \preceq 
    O(L^2 \rho) \cdot \LL_G^{+/2}\LL_C \LL_G^{+/2}$.
\end{lemma}
\begin{proof}
    We prove the first inequality. 
    The other one follows by a similar argument.
    Since
    \begin{align*}
        \norm{ \LL_G^{+/2} \Ltil \LL_G^{+/2} x }^2 
        &\leq 
        \paren{\norm{ \LL_G^{+/2} \LL_{\vec{F}} \LL_G^{+/2} x }
        + \norm{ \LL_G^{+/2} \LL_S \LL_G^{+/2} x }}^2
        \\
        &\leq 
        2 \norm{\LL_G^{+/2} \LL_{\vec{F}} \LL_G^{+/2} x }^2 + 
        2 \norm{ \LL_G^{+/2} \LL_S \LL_G^{+/2} x }^2
    \end{align*}
    for any conforming vector $x$, we can decompose the LHS into two terms 
    \[
        \LL_G^{+/2} (\Ltil^\top \LL_G^+ \Ltil) \LL_G^{+/2}
        \preceq
        2 \LL_G^{+/2} (\LL_{\vec{F}}^\top \LL_G^+ \LL_{\vec{F}}) \LL_G^{+/2}
        +2 \LL_G^{+/2} (\LL_S^\top \LL_G^+ \LL_S) \LL_G^{+/2}
    \]
    By \Cref{lemma:cyclefpart}, the first term is bounded by 
    \[
        \LL_G^{+/2} (\LL_{\vec{F}}^\top \LL_G^+ \LL_{\vec{F}}) \LL_G^{+/2}
        \preceq O(L^2 \rho) \II.
    \]
    For the second term, we first rearrange the terms from
    \Cref{lemma:cyclespart} to get
    \[
        \LL_S^{1/2} \LL_G^+ \LL_S^{1/2} \preceq L \rho \II.
    \]
    By multiplying appropriate terms on both sides and by \Cref{fact:twoside},
    \[
        \LL_G^{+/2} \LL_S \LL_G^+ \LL_S \LL_G^{+/2}
        \preceq
        O(L\rho) \LL_G^{+/2} \LL_S \LL_G^{+/2}
        \preceq
        O(L\rho) \LL_G^{+/2} \LL_C \LL_G^{+/2},
    \]
    which completes the proof.
\end{proof}

We now recall the Matrix Bernstein inequality which will be used to analyze the error involved in our random construction of sparsifiers.

\begin{theorem}[\cite{Tropp12} Matrix Bernstein] \label{thm:bernstein}
    Let $\XX_1, \ldots, \XX_m \in \mathbb{R}^{n_1 \times n_2}$ be independent
    random matrices that satisfies for any $i$, $av \XX_i = 0$ and $\| \XX_i \|
    \leq R$.
    Let the matrix variance be $\sigma^2 = 
    \max \curlyparen{ \norm{ \sum_{i=1}^m \av \XX_i^\top \XX_i }, 
    \norm{ \sum_{i=1}^m\av \XX_i \XX_i^\top } }$.
    Then, for all $\epsilon > 0$,
    \[
        \pr\left( \norm{ \sum_{i=1}^m \XX_i } \geq \epsilon \right) 
        \leq (n_1 + n_2) \cdot
        \exp\left( \frac{-\epsilon^2/2}{\sigma^2+R\epsilon/3} \right).
    \]
\end{theorem}

With now Matrix Bernstein and the previous helper lemmas, we state the sparsification and error analysis of running \CSO (which involves a random selection of what edges to preserve in a cycle).

\begin{lemma} \label{lemma:cyclekey}
    Given a directed Eulerian graph $\vec{G}$ whose edge weights are integral powers of
    2, and additionally 2-approximate effective resistances $\rr$ in $G$, 
    the algorithm \CSO returns a directed Eulerian graph $\vec{H}$ with edge
    weights still being powers of 2 such that if the number of edges in $G$
    satisfy $m = \Omega(\hat{m}\log n + nL^2 \log n)$, then with high probability,
    the number of edges in $\vec{H}$ is at most $\frac{15}{16}m$ and
    \[
        \norm{ \LL_G^{+/2} (\LL_{\vec{G}} - \LL_{\vec{H}}) \LL_G^{+/2} } \leq 
        O\left( \sqrt{\frac{nL^2 \log n}{m}} \right).
    \]
    The algorithm runs in $O(m) + \tcd(m,n)$ time.
\end{lemma}
\begin{proof}
    Recall that whether a cycle $\vec{C}$ is sampled as clockwise or
    counter-clockwise, the difference between the in-degrees and out-degrees of
    the vertices does not change.
    Thus, the final $\vec{H}$ is Eulerian after all the updates in
    \autoref{alg:cycleonce_add}. 

    Since $\rr$ is a 2-approximate effective resistances, 
    it suffices to take $\rho = O(n/m)$ so that
    each edge $e \in G$ satisfies
    $\lev_G(e) \leq \rho$ after the first step.
    Let $\XX_{i,j}$ be the matrix random variable corresponding to cycle
    $C_{i,j}$ such that 
    \[
        \begin{cases}
            \LL_G^{+/2} \Ltil_{i,j} \LL_G^{+/2}   & \mbox{w.p. } 1/2 \\
            -\LL_G^{+/2} \Ltil_{i,j} \LL_G^{+/2} & \mbox{w.p. } 1/2\\
        \end{cases}.
    \]
    Recall from \eqref{eq:diff}, this captures the changes on directed graph
    Laplacian incurred by \autoref{alg:cycleonce_add}.

    For any valid pair of $i,j$, $\| \XX_{i,j} \| = 
    \| \LL_G^{+/2} \Ltil_{i,j} \LL_G^{+/2}\| \leq 
    \sqrt{O(L^2\rho) \| \LL_G^{+/2} \LL_{C_{i,j}} \LL_G^{+/2} \|} = O(L^2\rho)$ by
    \Cref{lemma:cyclebounds} and the restriction of edges with at most
    $\rho$ leverage score.

    For the variance term, consider first 
    $\norm{ \sum_{i,j} \av \XX_{i,j}^* \XX_{i,j} }$.
    Now, $\av \XX_{i,j}^* \XX_{i,j} = 
    \LL_G^{+/2} \Ltil_{i,j}^\top \LL_G^+ \Ltil_{i,j} \LL_G^{+/2} 
    \preceq O(L^2\rho) \LL_G^{+/2}\LL_{C_{i,j}}
    \LL_G^{+/2}$ by \Cref{lemma:cyclebounds}.
    Then,
    \begin{align*}
        \sum_{i,j} \av \XX_{i,j}^* \XX_{i,j} 
        &\preceq
        O(L^2\rho) \cdot \sum_{i,j} \LL_G^{+/2}\LL_{C_{i,j}} \LL_G^{+/2}
        \\
        &\preceq
        O(L^2\rho) \cdot 
        \LL_G^{+/2} \left(\sum_{i,j} \LL_{C_{i,j}} \right) \LL_G^{+/2}
        \\
        &\preceq 
        O(L^2 \rho) \cdot \II
    \end{align*}
    by \Cref{lemma:cyclebounds} and the fact for each $j$ all cycles from $G_i$
    are edge disjoint.
    Thus, $\| \sum_{i,j} \av \XX_{i,j}^* \XX_{i,j} \| = O(L^2 \rho)$.
    Similarly, $\| \sum_{i,j} \av \XX_{i,j} \XX_{i,j}^* \| = O(L^2 \rho)$.

    We can apply \Cref{thm:bernstein} with $R=O(L^2 \rho)$ and 
    $\sigma^2 = O(L^2 \rho)$.
    Since $m = \Omega(nL^2 \log n)$, with probability $1- n^{-O(1)}$, we get the
    desired error bound 
    \[
        \norm{\LL_G^{+/2} (\LL_{\vec{G}} - \LL_{\vec{H}}) \LL_G^{+/2}} \leq 
        O\left( \sqrt{\frac{nL^2 \log n}{m}} \right).
    \]

    Consider the number of edges in $\vec{H}$. 
    Since $\rr$ is 2-approximate effective resistances, $\sum_e w_e r_e \leq
    2\sum_e \lev_G(e) \leq 2(n-1)$.
    The total number of edges with $w_e r_e \geq \frac{4n}{m}$ is at most
    $\frac{m}{2}$.
    Since we apply short cycle decomposition for $s=O(\log n)$ graphs, at most
    $O(n\log n + \hat{m} \log n) \leq \frac{m}{4}$ are not in any cycles
    if we choose $m = \Omega(\hat{m} \log n + n \log n)$ for some appropriate
    constant.
    Thus, at least $\frac{m}{4}$ edges are in the cycles.
    The expected fraction of edges that are added by \autoref{alg:cycleonce_add}
    is $\frac{1}{2}$.
    As there are at least $\frac{m}{4L}$ cycles and the length of each cycle is
    bounded by $L$, as long as $L = n^{o(1)}$ and $m = \Omega(n)$, by a Chernoff
    bound, at most $\frac{3}{4}$ fraction of the cycle edges are added to
    $\vec{H}$ with high probability, giving us at least $\frac{1}{16}$ fraction
    of the edges removed as required.

    Now, consider the runtime of the algorithm. 
    Since the number of edges across $G_1, \ldots, G_s$ is $O(m)$, the
    runtime except short cycle decompositions is $O(m)$ as well.
    Due to the super-additivity assumption of $\tcd(\cdot,n)$, the runtime for
    short cycle decompositions is bounded by $\tcd(m,n)$.
\end{proof}

We now provide the guarantees of \CS, which repeatedly calls \CSO until a
criterion on the number of edges is met.

\begin{theorem} \label{thm:cycledecomp}
    Given as input an Eulerian graph $\vec{G}$ with polynomial bounded integral edge
    weights and $\epsilon \in (0,1/2)$,
    the algorithm \CS returns a Eulerian graph $\vec{H}$ with 
    $O(\hat{m}\log n + \epsilon^{-2} n L^2 \log n)$ edges such that with high
    probability,
    \[
        \norm{ \LL_G^{+/2} (\LL_{\vec{G}} - \LL_{\vec{H}}) \LL_G^{+/2} } \leq
        \epsilon.
    \]
    The algorithm runs in time $O(m \log^2 n) + \tcd(O(m\log n), n)$.
\end{theorem}
\begin{proof}
    By \Cref{lemma:cyclekey}, the number of edges reduces by a constant factor
    each iteration with high probability.
    By a union bound, this holds over $O(\log(m\log n/ n)) = O(\log n)$
    iterations with high probability, which also gives the total number of
    iterations.

    By \Cref{lemma:cyclekey}, this geometric decrease in number of edges implies
    that the total error over all iterations is bounded up to a constant factor
    by the error in the last round, assuming the desired error bound as in
    \Cref{lemma:cyclekey}:
    \[
        O\left( \sqrt{\frac{nL^2 \log n}{m}} \right)
    \]
    where $m$ here is the number of edges in the last round.
    Note that since each approximation error bound in \Cref{lemma:cyclekey} with
    high probability, the error bound above also holds with high probability
    over $O(\log n)$ iterations.
    By the stopping condition of $m = \Omega(n\epsilon^{-2} L^2 \log n)$, with an
    appropriate constant we get the final error of at most $\epsilon$.
    This small error also implies that our 1.5-approximate effective resistances
    $\rr$ stays as 2-approximate throughout the algorithm.

    Consider the runtime of the algorithm.
    \Cref{lemma:cyclekey} gives a runtime of
    \begin{align*}
        & O\left(\sum_{i=1}^{O(\log n)} \left(\frac{15}{16} \right)^i m \right) 
        + 
        \sum_{i=1}^{O(\log n)} 
        \tcd\left( \left(\frac{15}{16} \right)^i m, n \right)
        \\
        \leq&
        O\left(\sum_{i=1}^{O(\log n)} \left(\frac{15}{16} \right)^i m \right) + 
        \tcd\left( \sum_{i=1}^{O(\log n)} 
        \left(\frac{15}{16} \right)^i m\log n, n \right)
        \\
        =& 
        O(m) + \tcd(O(m \log n),n)
    \end{align*}
    with high probability, where the first inequality holds by the
    super-additivity assumption of $\tcd$.
    Combine with a one time overhead of $O(m\log^2 n)$ \cite{KLP12} for
    computing the approximate effective resistances, we get the final runtime
    bound of $O(m\log^2 n) + \tcd(O(m \log n), n)$.
\end{proof}

Plugging in \Cref{lemma:cdfast}, we obtain the improved results on constructing
Eulerian Sparsifiers with short cycle decompositions, summarized in
\Cref{thm:informalcycle}.
\informalcycle*

\section{Sparsification via partial colouring}
\label{sec:existential}

In the previous algorithm \CS, the approach to sparsifying was to randomly pick
one part of each cycle (out of a partitioning of the cycle into two parts) to
remove from the graph.
The analysis then followed by observing on average this leads to a good
approximation, and that furthermore the variance in this random construction is
sufficiently small.
In this section, we show, however, that by using recent partial colouring results on
operator norm discrepancy bodies to pick what parts of a cycle to remove, we can
obtain better sparsifiers.

The main partial colouring result we use, relevant for picking a subset of
matrices to keep with minimal error, is restated below.

\colouring*

For this section, we assume the short cycle decomposition guarantees by
\Cref{lemma:cdfast} with $\hat{m} = O(n\log n)$ and $L = O(\log n)$.

\begin{algorithm}[t]
    \caption{Sparsification via partial colouring}
    \label{alg:coloursparse}
    \Pro{\PCS{$\vec{G}, \epsilon$}}{
        Decompose each edge by its binary representation.\;
        Compute $\rr$ a 1.5-approximate effective resistances in $G$.\;
        Let $\ol{S}$ be a set of cycles initialized to empty and let $\ol{x}$
        be its corresponding partial colouring.\;
        Set $\vec{G}' \gets \vec{G} +$ \CW{$\ol{S},\ol{x}$}.\;
        \While{$m' \geq O(n \epsilon^{-2}\log^2 n (\log\log n)^2 + 
        \epsilon^{-4/3} n \log^{8/3} n)$}{
            \If{$4m \geq m'$}{
                $\vec{G},\vec{G}',\ol{S},\ol{x} \gets$ 
                \PCG{$\vec{G},\vec{G}',\ol{S},\ol{x},\rr$}.\;
            }\Else{
                $\vec{G},\vec{G}',\ol{S},\ol{x} \gets$ 
                \PCC{$\vec{G},\vec{G}',\ol{S},\ol{x},\rr$}.\;
            }
        }
        \Return{$\vec{G}'$.}\;
    }
\end{algorithm}

\begin{algorithm}[t]
    \caption{Reweight a set of cycles based on colouring}
    \label{alg:cweight}
    \Pro{\CW{$S,x$}}{
        Let $\vec{H}$ be an empty directed graph.\;
        \For{each cycle $C \in S$ and corresponding directed cycle
        $\vec{C}$}{
            Add all the clockwise (resp. counter-clockwise) edges in $\vec{C}$
            with $1+x_C$ (resp. $1-x_C$) times their weight to $\vec{H}$.
            Note if $1+x_C = 0$ (resp. $1-x_C = 0$) the corresponding edge is
            not added.\;
            \label{alg:cw_add}
        }
        \Return{$\vec{H}$.}\;
    }
\end{algorithm}

For each cycle $C$ with its corresponding directed cycle $\vec{C}$, we set
$\AA(C) = \hlift(\LL_{G'}^{+/2} (\LL_{\vec{F}_C} - \LL_{S_C}) \LL_{G'}^{+/2})$ where
$\vec{F}_C$ is the cycle $\vec{C}$ with all edges set in clockwise direction and
$S_C$ is undirected graph with the set of edges corresponding to the
counter-clockwise edges in $\vec{C}$, same as in \Cref{sec:cycle}.
Note that this orientation is set initialy by \CO after a short cycle
decomposition step and fixed through out the execution.
Given a set of cycles $S$, we let $\calA[S]$ be the collection $\{ \AA(C) \}_{C
\in S}$.

\begin{algorithm}[t]
    \caption{Sparsify once for the $\vec{G}$ portion of $\vec{G}'$}
    \label{alg:colourg}
    \Pro{\PCG{$\vec{G},\vec{G}',\ol{S},\ol{\xx},\rr$}}{
        \KwIn{A directed Eulerian graph $\vec{G}$ where edge weights are
            integral powers of 2,%
            a set of cycles $\ol{S}$ where each cycle is edge disjoint from
            $G$,%
            a partial colouring $\ol{\xx} \in (-1,1)^{\ol{S}}$,%
            a graph $\vec{G}' = \vec{G} +$ \CW{$\ol{S},\ol{\xx}$},%
            a 2-approximate effective resistances $\rr$ in $G'$.}
        \KwOut{A directed Eulerian graph $\vec{H}$ where edge weights are
            integral powers of 2,%
            a set of cycles $\ol{T}$ where each cycle is edge disjoint from
            $H$,%
            a partial colouring $\ol{\zz} \in (-1,1)^{\ol{T}}$,%
            a graph $\vec{H}' = \vec{H} +$ \CW{$\ol{T},\ol{z}$}.}
        \BlankLine
        Let $\vec{H} \gets \vec{G}$ with only the edges which satisfy $w_e r_e
        > \frac{16n}{m'}$ and remove them from $\vec{G}$.\;
        \label{alg:colourg_lev}
        Partition $\vec{G}$ into uniformly weighted graph 
        $\vec{G}_1, \ldots, \vec{G}_q$ where $\vec{G}_i$ has all edge weights
        $2^i$ and $q = O(\log n)$.\;
        Let $S$ be the set of all cycles after applying \CD on $\vec{G}_1,
        \ldots, \vec{G}_s$ and set
        $\vec{H} \gets \vec{H} + \sum_{i=1}^s \vec{G}_i \backslash 
        \left( \bigcup_{j=1}^t \vec{C}_{i,j} \right)$.\;
        $T', \ol{T}',\yy,\ol{\yy} \gets$ \CT{$S, \vzero, \frac{1}{8}m$}.\;
        If \CW($T',\yy$) has more edges than \CW($T',-\yy$), we take $\yy
        \gets -\yy$ and $\ol{\yy} \gets -\ol{\yy}$.\;
        \label{alg:colourg_flip}
        $\vec{H} \gets \vec{H} + \CW(T',\yy)$.\;
        \label{alg:colourg_first}
        $\ol{T} \gets \ol{T}' \cup \ol{S}$ and set 
        $\ol{\zz} \gets \ol{\yy} + \ol{\xx}$.\;
        \label{alg:colourg_second}
        $\vec{H}' \gets \vec{H} +$ \CW{$\ol{T},\ol{\zz}$}.\;
        \Return{$\vec{H},\vec{H}',\ol{T},\ol{\zz}$.}\;
    }
\end{algorithm}

\begin{algorithm}[t]
    \caption{Sparsify once for the $\ol{S}$ portion of $\vec{G}'$}
    \label{alg:colours}
    \Pro{\PCC{$\vec{G},\vec{G}',\ol{S},\ol{\xx},\rr$}}{
        \KwIn{A directed Eulerian graph $\vec{G}$ where edge weights are
            integral powers of 2,%
            a set of cycles $\ol{S}$ where each cycle is edge disjoint from
            $G$,%
            a partial colouring $\ol{\xx} \in (-1,1)^{\ol{S}}$,%
            a graph $\vec{G}' = \vec{G} +$ \CW{$\ol{S},\ol{\xx}$},%
            a 2-approximate effective resistances $\rr$ in $G'$.}
        \KwOut{A directed Eulerian graph $\vec{H}$ where edge weights are
            integral powers of 2,%
            a set of cycles $\ol{T}$ where each cycle is edge disjoint from
            $H$,%
            a partial colouring $\ol{\zz} \in (-1,1)^{\ol{T}}$,%
            a graph $\vec{H}' = \vec{H} +$ \CW{$\ol{T},\ol{z}$}.}
        \BlankLine
        Set $\ol{S}'$ be an empty set of cycles initialy.
        For each $C \in \ol{S}$, let $C'$ be $C$ with its weight by
        $(1-|\ol{x}_C|)$ and add $C'$ to $\ol{S}'$.\;
        $T',\ol{T}',\yy,\ol{\yy} \gets$ \CT{$\ol{S}',\vzero,\frac{1}{4}m'$}\;
        \If{$m(\{C' \in T' : |\ol{x}_C - (1-|\ol{x}_C|) y_{C'}| = 1 \}) > 
            m(\{C' \in T' : |\ol{x}_C + (1-|\ol{x}_C|) y_{C'}| = 1 \})$}{
            $\yy \gets -\yy, \ol{\yy} \gets -\ol{\yy}$.\;
            \label{alg:colours_flip}
        }
        Set $\zz, \ol{\zz}$ to be the parts of 
        $\ol{\xx}+(1-|\ol{\xx}|) \circ (\yy+\ol{\yy})$ 
        with magnitude $1$ and $<1$ respectively.
        Here we abused $\circ$ to let $C$ and $C'$ refering to the same index,
        Set the partition $T,\ol{T}$ of $\ol{S}$ accordingly.
        \;
        \label{alg:colours_first}
        $\vec{H} \gets \vec{H} + \CW(T,\zz)$.\;
        $\vec{H}' \gets \vec{H} +$ \CW{\ol{T},\ol{\zz}}.\;
        \Return{$\vec{H},\vec{H}',\ol{T},\ol{\zz}$.}\;
    }
\end{algorithm}

\begin{algorithm}[t]
    \caption{Partial colouring cycles with target mass}
    \label{alg:colourtarget}
    \Pro{\CT{$S,y,m_t$}}{
        \KwIn{A set of cycles $S$ of size $s = |S|$, \\
            a partial colouring $y \in (-1,1)^S$,\\ 
            and a target mass of $m_t$ edges.}
        \KwOut{A set of fully coloured cycles $S \backslash \ol{S}$ with colouring
            $x$, \\
            A set of partially coloured cycles $\ol{S}$ with colouring $\ol{x}$
            satisfying $\ol{x} \in (-1,1)^{\ol{S}}$.} 
        \BlankLine
        Initialize $x = 0$ be a empty colouring over $S$.\;
        Define $\ol{S}$ to always be the set of non-fully coloured cycles in
        $S$ and let $\ol{s} = |\ol{S}|$ always. 
        Set $\ol{x}$ be the partial colour on $\ol{S}$ always.\;
        \While{$\ol{s} > \frac{m_t}{L}$}{
            $x[\ol{S}] \gets$ \PC{$\calA[\ol{S}],\ol{x}$}.\;
        }
        Let $\ol{x} \gets x$ with entries of magnitude $<1$ and set 
        $x \gets x-\ol{x}$.\;
        \Return{$S \backslash \ol{S}, \ol{S}, x, \ol{x}$.}\;
    }
\end{algorithm}

\CW is our partial colouring alternative of the random selection of edges in a
cycle in \CSO.
It similarly does not change the difference between the in-degree and out-degree
and preserves integral weights, stated in Lemma~\ref{lemma:cwdeg}.

\begin{lemma} \label{lemma:cwdeg} %
    Given a set of cycles $S$ where each cycle is uniformly weighted,
    and any partial colouring $x \in [-1,1]^{S}$, the
    algorithm \CW returns a directed graph $\vec{H}$ such that the difference in
    the in and out degrees are the same as in 
    $\sum_{C \in S} \vec{C}$.

    In addition, if $x \in \{ \pm 1\}^S$, $\vec{H}$ also has integral edge
    weights with the largest edge weight at most twice the largest edge weight
    in cycles in $S$.
\end{lemma}
\begin{proof}
    For the degree condition, it suffices to consider a single cycle $C$ and
    show that the reweighted directed cycle, say $\vec{C}'$ in
    \autoref{alg:cw_add} preserves the differences of the in and out degrees of
    $\vec{C}$.
    Recall the definition of $\vec{F}$ and $S$ of $C$, see \CO in
    \Cref{alg:correct_orientation}, and the argument in \Cref{sec:cycle} for
    showing degree differences preservation under the special case of 
    $x \in \{ \pm 1 \}$.
    Note first that the edge weights are the same.
    Either a vertex had an incoming and outgoing edge (and so difference $0$),
    in which case both edges are either in $\vec{F} \setminus S$ or in $S$ and
    hence always added together with the same weights of (so still difference
    $0$).
    Alternatively a vertex has two incoming or outgoing edges, in which case
    one edge gets a new weight of $1+x$ and the other gets $1-x$, which then
    still preserves the difference between in and out degree.

    If $x \in \{ \pm 1 \}$ the edge weights of $\vec{C}'$ is exactly twice that
    of $C$ unless $\vec{C}'$ is emtpy.
    Thus, $\vec{H}$ still has integral edge weights with largest weight at most
    doubled.
\end{proof}

For the rest of this section, we refer to a set of uniformly weighted cycles
(two cycles can have different weights) as a set of cycles for simplicity.
We write $m(S) = \sum_{C \in S} |E(C)|$ as the total number of edges in $S$.
In \PCS, \PCG and \PCC, by applying $'$ to a graph we mean $\vec{G}' = \vec{G}
+$ \CW{$\ol{S},\ol{x}$}.
We denote $m'$ as the number of edges in $\vec{G}'$.
Note that this is the primary number of edges we consider rather than $m$.

Towards analyzing \PCS, we first state the guarantees of the \CT subroutine
which guarantees a partial colouring of at least a specified size.
\begin{lemma} \label{lemma:colourtarget}
    The returned values of \CT{$S,y,m_t$} satisfy that 
    $m(\ol{S}) \leq m_t$ and the number of calls to \PC is
    $O(\log(|S|L/m_t))$.
    
    In addition, 
    If the set of cycles $S$ satisfies 
    $\sum_{C \in S} \| \AA(C) \| \leq \sigma^2$ and
    $\sum_{C \in S} \| \AA(C) \|_F^2 \leq v$, then the output of \CT{$S,y,m_t$}
    satisfies that
    \[
        \norm{ \sum_{C \in S}  (x+\ol{x} - y) \AA(C) } \leq 
        O\left(\sigma \cdot \log\left(\frac{|S|L}{m_t} \right) + 
        (\log^{3/4} n) \sigma^{1/2} \left(\frac{vL}{m_t} \right)^{1/4} \right)
    \]
\end{lemma}
\begin{proof}
    Note that each cycle has its number of edges bounded by $L$, hence we have
    $m(\ol{S}) \leq L |\ol{S}| \leq m_t$ by the terminating condition of
    the while loop in \CT.

    For the number of calls, note at each call to the while loop, we have the
    size of $\ol{S}$ gets decreased by a factor of $1-c'$ where $c'$ is the
    universal constant in \PC.
    Hence by the ith round we have $|\ol{S}| \leq (1-c')^{i}|S|$ and at
    termination this is $\leq \frac{m_t}{L}$.
    So we have the number of iterations is the smallest $i$ such that 
    $(1-c')^{i}|S| \leq \frac{m_t}{L}$.
    Rearranging we get $i = O(\log(\frac{|S|L}{m_t}))$, showing the
    claimed bound on the number of iterations.

    Consider the error bound.
    Combine the number of iterations with the first term in \eqref{eq:colour}
    of \Cref{lemma:colouring}, we get our desired first term. 
    For the second term, recall from above that $|\ol{S}|$ decreases
    geometrically.
    Then $f = (v / |\ol{S}|)^{1/2}$ increases exponentially over the iterations.
    Hence the sum of the second terms in \eqref{eq:colour} is bounded by the
    last one with $f = O((vL / m_t)^{1/2})$, giving us
    \[
        O((\log^{3/4} n) \sigma^{1/2} f^{1/2}) = 
        O((\log^{3/4} n) \sigma^{1/2} (vL/m_t)^{1/4})
    \]
    as required.
\end{proof}

With now the guarantees for the subroutines \CT and \CW, we analyze \PCG and
\PCC which use them.

\begin{lemma} \label{lemma:gerror}
    If the input graphs $\vec{G},\vec{G}'$ satisfy $4m \geq m'$ and 
    the input set of cycles $\ol{S}$ and it corresponding partial colours
    $\ol{x}$ satisfies that each cycle $C \in \ol{S}$ has $w_e r_e \leq
    \frac{4n}{m'}$ for each edge $e \in C$,
    the algorithm \PCG returns $\vec{H}$ with edge weights
    still being powers of 2 and at most twice the largest weight in $\vec{G}$,
    a set of cycles $\ol{T}$ with its corresponding partial colours $\ol{\zz}$
    satisfying $\vec{H}' = \vec{H} +$ \CW{$\ol{T},\ol{\zz}$} is an Eulerian
    graph and each cycle $C \in \ol{T}$ also has $w_e r_e \leq \frac{4n}{m_H'}$
    for each edge $e \in C$, where $m_H' = |E(\vec{H})|$.
    and,
    \[
        \norm{ \LL_{G'}^{+/2} (\LL_{\vec{G}'} - \LL_{\vec{H}'}) \LL_{G'}^{+/2}}
        \leq 
        O\left( \sqrt{\frac{n\log^2 n}{m'}} \log \log n + 
        \left(\frac{n \log^{8/3} n}{m'} \right)^{3/4} \right).
    \]
\end{lemma}
\begin{proof}
    The edge weights condition of $\vec{H}$ is guaranteed by
    \Cref{lemma:cwdeg}.
    Also by \Cref{lemma:cwdeg}, both $\vec{H}$ and $\vec{H}'$ are Eulerian.

    We now show first that the output cycles $\ol{T}$ still satisfy the
    approximate leverage score condition.
    To do so, we, in fact, prove that every cycle $C$ arising throughout the
    algorithm satisfies that $w_e r_e \leq \frac{4n}{m'}$.
    Note first that $m_H' < m'$ always.
    Each cycle can only originate from one of the two sets: $S$ and $\ol{S}$.
    The condition for cycles in $\ol{S}$ follows by assumption. 
    For cycles in $S$, the first line guarantees the condition as well.

    Since $\rr$ is a 2-approximate effective resistances, 
    it suffices to take $\rho = O(n/m')$ to ensure 
    $\lev_{G'}(e) \leq \rho$ after the first step.
    By \autoref{alg:colourg_first} and \autoref{alg:colourg_second},
    the output Eulerian graph $\vec{H}'$ satisfies that
    \[
        \hlift\left( \LL_{G'}^{+/2} 
        (\LL_{\vec{H}'} - \LL_{\vec{G}'}) \LL_{G'}^{+/2} \right)
        =
        \sum_{C \in S} (y'_C + \ol{y}'_C - 0) \AA(C)
    \]
    where all vectors are taken as the final values in an execution.

    By definition of Hermitian lift, each matrix $\AA(C)^2$ is block diagonal
    with blocks $\LL_{G'}^{+/2} \Ltil_{\vec{C}}^\top \LL_{G'}^+ \Ltil_{\vec{C}}
    \LL_{G'}^{+/2}$ and $\LL_{G'}^{+/2} \Ltil_{\vec{C}} \LL_{G'}^+
    \Ltil_{\vec{C}}^\top \LL_{G'}^{+/2}$. 
    Here $\Ltil_{\vec{C}} = \LL_{\vec{F}} - \LL_{S}$ with fixed orientation 
    (recall \CO).
    Since every cycle $C \in S$ satisfies $\lev_{G'}(e) \leq
    \rho$ for each $e \in C$,
    by \Cref{lemma:cyclebounds}, both matrices are spectrally bounded above by 
    $O(L^2 \rho) \cdot \LL_{G'}^{+/2} \LL_C \LL_{G'}^{+/2}$.
    Thus, as $G$ is a subgraph of $G'$,
    \[
        \sum_{C \in S} \AA(C)^2 
        \preceq 
        O(L^2 \rho) \cdot 
        \begin{bmatrix}
            \LL_{G'}^{+/2} \left(\displaystyle\sum_{C \in S} \LL_C
            \right) \LL_{G'}^{+/2} & \\
            & \LL_{G'}^{+/2} \left(\displaystyle\sum_{C \in S} \LL_C
            \right) \LL_{G'}^{+/2} \\
        \end{bmatrix}
        \preceq O(L^2 \rho)
        \begin{bmatrix}
            \II & \\
            & \II \\
        \end{bmatrix}.
    \]
    The sum of Frobenius norm squared is bounded by
    \[
        \sum_{C \in S} \| \AA(C) \|_F^2 
        =
        \sum_{C \in S} \tr{\AA(C)^2} 
        =
        \tr{ \sum_{C \in S} \AA(C)^2 }
        \leq 
        \tr{ O(L^2 \rho) 2\II}
        = O(nL^2 \rho).
    \]
    We can now apply \Cref{lemma:colourtarget} with 
    $m_t = \frac{1}{8}m , \sigma^2 = O(L^2 \rho)$ and $v = O(nL^2 \rho)$ to get
    \begin{align*}
        \norm{ \sum_{C \in S} (y'_C + \ol{y}'_C - 0) \AA(C) }
        &\leq
        O\left(\sqrt{L^2 \rho} \cdot \log\left(\frac{8|S|L}{m} \right) + 
        (\log^{3/4} n) (L^2 \rho)^{1/4} 
        \left(\frac{8nL^3 \rho}{m} \right)^{1/4} \right)
        \\
        &=
        O\left( \sqrt{\frac{nL^2}{m'}} \log L + 
        \left(\frac{nL^{5/3} \log n}{m'} \right)^{3/4} \right)
    \end{align*}
    where we used $|S| \leq m$ and $m = \Theta(m')$.
    Finally, note that
    \[
        \norm{\LL_{G'}^{+/2} (\LL_{\vec{G}'} - \LL_{\vec{H}'}) \LL_{G'}^{+/2}}
        =
        \norm{ \hlift\left( \LL_{G'}^{+/2} 
        (\LL_{\vec{G}'} - \LL_{\vec{H}'}) \LL_{G'}^{+/2} \right) }
        =
        \norm{ \sum_{C \in S} (y'_C + \ol{y}'_C - 0) \AA(C) }.
    \]
\end{proof}

Before we prove the approximation guarantees for \PCC, we show the following
lemma regarding scaling matrices in the set of extra cycles $\ol{S}$. 
\begin{lemma} \label{lemma:scale}
    For directed Eulerian graph $\vec{G}$, a set of cycles $\ol{S}$ where each
    cycle $C \in \ol{S}$ satisfies that $\vec{G}$ and $\vec{C}$, the
    corresponding directed cycle of $C$, are edge-disjoint.
    Let $\ol{x} \in (-1,1)^{S}$ be a partial colouring on $\ol{S}$.
    Then the directed Eulerian graph $\vec{G}' = \vec{G} +$ \CW{$\ol{S},\ol{x}$}
    satisfies 
    \[
        \LL_G + \sum_{C \in \ol{S}} (1-|\ol{x}_C|) \LL_C \preceq \LL_{G'}
    \]
\end{lemma}
\begin{proof}
    For any $C \in \ol{S}$, let $\vec{C'} =$ \CW{$C, \ol{x}_C$} where we abused
    the definition of \CW to take in a single cycle instead of a set of cycles.
    Note that the undirectification $\LL_{G'} = \LL_G + \sum_{C \in
    \ol{S}} \LL_{C'}$.

    Since $|\ol{x}_C| < 1$, all edges in $C$ must be present in $C'$ as well and
    the minimum edge weight is at least $1-|\ol{x}_C|$ times the original
    uniform edge weights of $C$.
    Hence, since undirected Laplacians are PSD,
    \[
        (1-|\ol{x}_C|) \LL_C \preceq \LL_{C'}
    \]
    Summing over all $C$, we get
    \[
        \LL_G + \sum_{C \in \ol{S}} (1-|\ol{x}_C|) \LL_C 
        \preceq
        \LL_G + \sum_{C \in \ol{S}} \LL_{C'}
        =
        \LL_{G'}.
    \]
\end{proof}

\begin{lemma} \label{lemma:serror}
    If the input set of cycles $\ol{S}$ and it corresponding partial colours
    $\ol{x}$ satisfies that each cycle $C \in \ol{S}$ has $w_e r_e \leq
    \frac{4n}{m'}$ for each edge $e \in C$,
    the algorithm \PCC returns $\vec{H}$ with edge weights
    still being powers of 2 and at most twice the largest weight in $\vec{G}$,
    a set of cycles $\ol{T}$ with its corresponding partial colours $\ol{y}$
    satisfying $\vec{H}' = \vec{H} +$ \CW{$\ol{T},\ol{y}$} is an Eulerian graph
    and each cycle $C \in \ol{T}$ also has $w_e r_e \leq \frac{4n}{m_H'}$ for
    each edge $e \in C$, where $m_H' = |E(\vec{H})|$.
    and,
    \[
        \norm{ \LL_{G'}^{+/2} (\LL_{\vec{G}'} - \LL_{\vec{H}'}) \LL_{G'}^{+/2}}
        \leq 
        O\left( \sqrt{\frac{n\log^2 n}{m'}} \log \log n + 
        \left(\frac{n \log^{8/3} n}{m'} \right)^{3/4} \right).
    \]
\end{lemma}
\begin{proof}
    The proof follows similarly to that of \Cref{lemma:gerror}.

    Again, the edge weights condition of $\vec{H}$ is guaranteed by
    \Cref{lemma:cwdeg}.
    Also by \Cref{lemma:cwdeg}, both $\vec{H}$ and $\vec{H}'$ are Eulerian.

    Observe that $m_H' \leq m$ always, and $\ol{T} \subset \ol{S}$.
    Then, the output cycles still satisfy the approximate leverage score
    condition.

    By \autoref{alg:colours_first},
    the output Eulerian graph $\vec{H}'$ satisfies
    \[
        \hlift\left( \LL_{G'}^{+/2} 
        (\LL_{\vec{H}'} - \LL_{\vec{G}'}) \LL_{G'}^{+/2} \right)
        =
        \sum_{C \in \ol{S}} (z_C + \ol{z}_C - x_C) \AA(C) 
        =
        \sum_{C \in \ol{S}} (1-|x_C|)(y_C + \ol{y}_C) \AA(C) 
    \]
    where all vectors are taken as the final values in an execution.
    By our definition of $C'$, $\AA(C') = (1-|x_C|) \AA(C)$ and
    \[
        \sum_{C \in \ol{S}} (1-|x_C|)(y_C + \ol{y}_C) \AA(C) 
        =
        \sum_{C' \in \ol{S}'} (y_C + \ol{y}_C) \AA(C') 
    \]

    By definition of Hermitian lift, each matrix $\AA(C)^2$ is block diagonal
    with blocks $\LL_{G'}^{+/2} \Ltil_{\vec{C}}^\top \LL_{G'}^+ \Ltil_{\vec{C}}
    \LL_{G'}^{+/2}$ and $\LL_{G'}^{+/2} \Ltil_{\vec{C}} \LL_{G'}^+
    \Ltil_{\vec{C}}^\top \LL_{G'}^{+/2}$. 
    Here $\Ltil_{\vec{C}} = \LL_{\vec{F}} - \LL_{S}$ with fixed orientation 
    (recall \CO).
    Since every cycle $C \in \ol{S}$ satisfies $\lev_{G'}(e) \leq
    \rho$ for each $e \in C$,
    by \Cref{lemma:cyclebounds}, both matrices are spectrally bounded above by 
    $O(L\rho) \cdot \LL_{G'}^{+/2} \LL_C \LL_{G'}^{+/2}$.
    Thus, by the disjointness of $G$ and $\ol{S}$,
    \begin{align*}
        & \sum_{C' \in \ol{S}'} \AA(C')^2 
        \preceq
        \sum_{C \in \ol{S}} (1-|x_C|) \AA(C)^2 
        \\
        \preceq &
        O(L\rho) \cdot 
        \begin{bmatrix}
            \LL_{G'}^{+/2} \left(\displaystyle\sum_{C \in \ol{S}} (1-|x_C|)
            \LL_C \right) \LL_{G'}^{+/2} & \\
            & \LL_{G'}^{+/2} \left(\displaystyle\sum_{C \in \ol{S}} (1-|x_C|)
            \LL_C \right) \LL_{G'}^{+/2} \\
        \end{bmatrix}
        \preceq O(L^2 \rho)
        \begin{bmatrix}
            \II & \\
            & \II \\
        \end{bmatrix}.
    \end{align*}
    where we used the PSD property of $\AA(C)^2$ and the fact $1-|x_C| \leq 1$
    for the first inequality and \Cref{lemma:scale} for the second inequality.

    The sum of Frobenius norm squared is bounded by
    \[
        \sum_{C' \in \ol{S}'} \| \AA(C') \|_F^2 
        \leq
        \sum_{C \in \ol{S}} (1 - |x_C|) \tr{\AA(C)^2} 
        =
        \tr{ \sum_{C \in \ol{S}} (1- |x_C|) \AA(C)^2 }
        = O(nL^2 \rho)
    \]
    using the variance bound from above.

    We can now apply \Cref{lemma:colourtarget} with 
    $m_t = \frac{1}{4}m', \sigma^2 = O(L^2 \rho)$ and $v = O(nL^2 \rho)$ to get
    \begin{align*}
        \norm{ \sum_{C' \in \ol{S}'} (y_C + \ol{y}_C - 0) \AA(C') }
        &\leq
        O\left(\sqrt{L^2\rho} \cdot \log\left(\frac{4|\ol{S}|L}{m'} \right) + 
        (\log^{3/4} n) (L^2 \rho)^{1/4} 
        \left(\frac{4nL^3 \rho}{m'} \right)^{1/4} \right)
        \\
        &=
        O\left( \sqrt{\frac{nL^2}{m'}} \log L + 
        \left(\frac{nL^{5/3} \log n}{m'} \right)^{3/4} \right)
    \end{align*}
    where we used $|\ol{S}'| = |\ol{S}| \leq m'$.
    Finally, note that
    \[
        \norm{\LL_{G'}^{+/2} (\LL_{\vec{G}'} - \LL_{\vec{H}'}) \LL_{G'}^{+/2}}
        =
        \norm{ \hlift\left( \LL_{G'}^{+/2} 
        (\LL_{\vec{G}'} - \LL_{\vec{H}'}) \LL_{G'}^{+/2} \right) }
        =
        \norm{ \sum_{C' \in \ol{S}'} (y_C + \ol{y}_C - 0) \AA(C) }.
    \]
\end{proof}

The sparsification induced by \PCSO is conditional, and we state the condition
and sparisification induced in \Cref{lemma:sparsity1}. 
However, even when the condition in \Cref{lemma:sparsity1} is not met, we are
guaranteed each \PCC will geometrically make progress towards satisfying the
condition needed for \Cref{lemma:sparsity1}. 
The alternative is stated in \Cref{lemma:sparsity2}.

\begin{lemma} \label{lemma:sparsity1}
    For inputs $\vec{G}, \vec{G}', \ol{S}, \ol{x}, \rr$ to \PCG satisfying that
    $4m \geq m' \geq \Omega(n \log^2 n)$, 
    the outputs satisfy that the number of edges in $\vec{H}'$ is upperbounded
    by $m_H' \leq \frac{63}{64} m'$.
\end{lemma}
\begin{proof}
    Since $\rr$ is 2-approximate effective resistances, 
    $\sum_e w_e r_e \leq 2(n-1)$, we have at most $\frac{1}{8}m'
    \leq \frac{1}{2}m$ edges are removed from $\vec{G}$ 
    in \autoref{alg:colourg_lev}.

    Since $m \geq \frac{1}{4}m' = \Omega(n \log^2 n)$ and the number of edges not
    in any cycle is $\hat{m}q = O(n \log^2 n)$, by picking an appropriate
    constant in $\Omega(n \log^2 n)$, we can guarantee the total
    number of edges in all cycles satisfies $m(S) \geq \frac{1}{4}m$.
    \Cref{lemma:colourtarget} then guarantees $m(\ol{T}') \leq \frac{1}{8}m$
    and that $m(T') \geq \frac{1}{8} m$.

    Now, by \CW, the total number of edges in \CW{$T',\yy$} and \CW{$T',-\yy$}
    is exactly $m(T')$.
    Thus, \autoref{alg:colourg_flip} means at least $\frac{1}{2}m(T')
    \geq \frac{1}{16}m \geq \frac{1}{64}m'$ edges are removed in total as
    required.
\end{proof}

Combined with the lemma above, the following lemma (which states the
guarantee in the alternative case) tells us that in every two 
iterations, the number of edges must decrease by at least a constant
factor as the conditions needed for \Cref{lemma:sparsity1} will be satisfied. 

\begin{lemma} \label{lemma:sparsity2}
    If inputs $\vec{G}, \vec{G}', \ol{S}, \ol{x}, \rr$ to \PCC satisfies that
    $4m < m'$ 
    , then either the number of edges in $\vec{H}'$ decreases to $m_H' \leq
    \frac{63}{64} m'$, or the number of edges in $\vec{H}$ satisfies $4m_H \geq
    m_H'$.
\end{lemma}
\begin{proof}
    Suppose $m_H' > \frac{63}{64}m'$.
    By \Cref{lemma:colourtarget}, $m(\ol{T'}) \leq \frac{1}{8}m'$.

    Since $\yy \in \{ \pm 1 \}^{T'}$, we have
    $\{C' \in T' : |\ol{x}_C - (1-|\ol{x}_C|) y_{C'}| = 1 \} \cup 
    \{C' \in T' : |\ol{x}_C + (1-|\ol{x}_C|) y_{C'}| = 1 \} = T'$.
    Let the two sets above be $T'_1$ and $T'_2$, 
    Then, $m(T'_1) + m(T'_2) \geq m(T')$ \footnotemark.
    \footnotetext{Contrary to the proof of \Cref{lemma:sparsity1}, this is an
    inequality since magnitude of 1 can be achieve using both $y_{C'}$ and
    $-y_{C'}$ if $x_C = 0$.}
    This means, after re-adjusting the colouring in \autoref{alg:colours_flip}, 
    \[
        m(\ol{T}) \leq \frac{1}{2}m(T') + m(\ol{T'}) 
        \leq \frac{1}{2}m' + \frac{1}{8}m'
        = \frac{5}{8}m' \leq \frac{40}{63} m_H'.
    \]
    Then,
    \[
        m_H = m_H' - m(\ol{T}) \geq \frac{23}{63} m_H' \geq \frac{1}{4} m_H'
    \]
    as required.
\end{proof}

With these analyses of \PCG and \PCC, we now state the guarantees of \PCS which
calls \PCG and \PCC until a desired sparisity is met and provides better
sparsifiers than \CS.
\begin{theorem}[\Cref{thm:informalexist} Formal] \label{thm:exist}
    Given input a Eulerian graph $\vec{G}$ with polynomial bounded integral edge
    weights and $\epsilon \in (0,1/2)$, 
    the algorithm \PCS returns in polynomial time a Eulerian graph $\vec{H}$
    with $O(n \epsilon^{-2} \log^2 n (\log\log n)^2 + 
    n \epsilon^{-3/4} \log^{8/3} n )$
    edges
    satisfying
    \[
        \norm{ \LL_G^{+/2} (\LL_{\vec{G}} - \LL_{\vec{H}}) \LL_G^{+/2} } \leq
        \epsilon.
    \]
\end{theorem}
\begin{proof}
    By \Cref{lemma:sparsity1} and \Cref{lemma:sparsity2}, in every two
    iterations the number of edges must decreases by at least a constant
    fraction, as the condition $4m \geq m'$ must be satisfied at least once. 
    Note that initialy $m = m' \geq \frac{1}{4} m'$ is satisfied.
    Thus, the total number of iterations is at most 
    $O(\log(m\log n / n)) = O(\log n)$ where the extra $\log n$ 
    comes from the decomposition by edge weights.

    By \Cref{lemma:gerror} and \Cref{lemma:serror}, the largest edge weight
    doubles each iteration.
    Thus, the edge weights in each $\vec{G}$ are still integral and polynomially
    bounded over $O(\log n)$ iterations.

    As the number of edges decreases geometrically every $O(1)$ iterations, the
    total error is bounded by constant factor times the error in the last round for
    both terms in \Cref{lemma:gerror} and \Cref{lemma:serror}:
    \[
        O\left( \sqrt{\frac{n\log^2 n}{m'}} \log \log n + 
        \left(\frac{n \log^{8/3} n}{m'} \right)^{3/4} \right).
    \]
    where $m'$ is the number of edges in $\vec{G}'$ in the last round.
    Since the algorithm stops at 
    $m' \geq \Omega(n\epsilon^{-2} \log^2 n (\log\log n)^2)$ and 
    $m' \geq \Omega(n \epsilon^{-3/4} \log^{8/3} n)$ edges, the largest of both
    terms must be bounded by $\frac{1}{2}\epsilon$ by picking appropriate
    constant for the stopping condition.
    
    This small error also implies that our 1.5-approximate effective resistances
    $\rr$ stays as 2-approximate throughout the algorithm.
    Then, by \Cref{lemma:gerror} and \Cref{lemma:serror}, the set of cycles
    $\ol{S}$ always satisfy $w_e r_e \leq \frac{4n}{m'}$ where $m'$ is the
    number of edges in $\vec{G}'$ throughout as required.

    \Cref{lemma:colouring} guarantees the polynomial running time of our
    algorithm.
\end{proof}

\section*{Acknowledgements}
This research was was supported by an NSERC Discovery Grant
RGPIN-2018-06398 and a Sloan Research Fellowship awarded to SS. AT is supported by a Vanier Fellowship from the Natural Sciences and Engineering Research Council of Canada.

We thank Arun Jambulapati for notifying us an issue in previous version of this
manuscript.

\printbibliography

\end{document}